\documentclass{mem}
\usepackage{natbib}\usepackage{txfonts}\usepackage{balance}
\usepackage{graphicx}
\usepackage[a4paper,breaklinks,dvipdfm]{hyperref}
\idline{75}{282}
\begin{document}
\def\kms{$\mathrm {km~s}^{-1}$}

\def\la{\;
\raise0.3ex\hbox{$<$\kern-0.75em\raise-1.1ex\hbox{$\sim$}}\; }
\def\ga{\;
\raise0.3ex\hbox{$>$\kern-0.75em\raise-1.1ex\hbox{$\sim$}}\; }

\newcommand{\nhhh}{NH$_3$}
\newcommand{\hcIII}{HC$_3$N}
\newcommand{\hcV}{HC$_5$N}
\newcommand{\hcVII}{HC$_7$N}
\newcommand{\dmm}{$\Delta\mu/\mu$}
\newcommand{\daa}{$\Delta\alpha/\alpha$}
\newcommand{\cmm}{cm$^{-3}$}

\title{
Limits on the space-time variations of fundamental constants
}

\author{
S. A. \,Levshakov\inst{1,2} 
\and 
C. \,Henkel\inst{3,4}
\and
D. \,Reimers\inst{5}
\and
P. \,Molaro\inst{6,7}
          }

\offprints{S. A. Levshakov}

\institute{
Ioffe Physical-Technical Institute, Russian Academy of Sciences,
Polytekhnicheskaya Str. 26, 194021 St.~Petersburg, Russia
\and
St.~Petersburg Electrotechnical University `LETI', Prof. Popov Str. 5,
197376 St.~Petersburg, Russia
\and
Max-Planck-Institut f\"ur Radioastronomie, Auf dem H\"ugel 69, D-53121 Bonn, Germany
\and
Astronomy Department, King Abdulaziz University, P.O.
Box 80203, Jeddah 21589, Saudi Arabia
\and
Hamburger Sternwarte, Universit\"at Hamburg,
Gojenbergsweg 112, D-21029 Hamburg, Germany
\and
INAF~-- Osservatorio Astronomico di Trieste, Via Tiepolo 11,
I-34131 Trieste, Italy
\and
Centro de Astrofísica, Universidade do Porto, Rua das Estrelas, 4150-762, Porto, Portugal
\email{lev@astro.ioffe.rssi.ru}
}

\authorrunning{Levshakov et al.}
\titlerunning{Limits on the space-time variations of fundamental constants}

\abstract{
We report on new tests that improve our previous (2009-2010) estimates 
of the electron-to-proton mass ratio variation, $\mu = m_{\rm e}/m_{\rm p}$.
Subsequent observations (2011-2013) at the Effelsberg 100-m telescope 
of a sample of eight molecular cores from the Milky Way disk
reveal systematic errors in the measured radial velocities 
varying with an amplitude $\approx \pm 0.01$ \kms\ during the exposure time.   
The averaged offset between the radial velocities of
the rotational transitions of
\hcIII(2-1), \hcV(9-8), \hcVII(16-15), \hcVII(21-20), and \hcVII(23-22),  
and
the inversion transition of \nhhh(1,1) 
gives
$\langle \Delta V \rangle = 0.002\pm0.015$ \kms\ [$3\sigma$ confidence level (C.L.)].
This value, when interpreted in terms of
\dmm\ $= (\mu_{\rm obs} - \mu_{\rm lab})/\mu_{\rm lab}$,
constraints the $\mu$-variation at the level of
$\Delta \mu/\mu  < 2\times10^{-8}$ ($3\sigma$ C.L.), 
which is the most stringent limit on the fractional changes in $\mu$
based on radio astronomical observations. 
}

\maketitle{}

\section{Introduction}

\begin{figure}[t!]
\vspace{0.0cm}
\resizebox{\hsize}{!}{\includegraphics[clip=true]{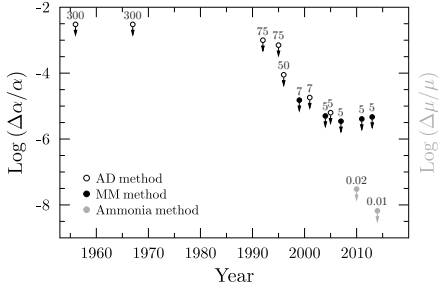}}
\vspace{-1.5cm}
\caption{\footnotesize
Astronomical constraints on $\alpha$- and $\mu$-variations ($1\sigma$ C.L.)
for the period from 1956 to 2013. Above each point, 
the spectral resolution (FWHM in \kms) is indicated.
$\alpha$-variation constraints are based on the alkali-doublet (AD)
and many-multiplet (MM) methods, 
whereas $\mu$-variation~-- on the
ammonia method. The data points for the AD method are taken from
Savedoff (1956), Bahcall et al. (1967),
Levshakov (1992), Varshalovich \& Potekhin (1995), Varshalovich et al. (1996),
Murphy et al. (2001), Chand et al. (2005);
MM method~-- Webb et al. (1999), 
Quast et al. (2004), Srianand et al. (2007), Agafonova et al. (2011), 
Molaro et al. (2013); 
ammonia method~-- from Levshakov et al. (2010a), Levshakov et al. (2013a). 
The figure shows that limits on the $\alpha$- and $\mu$-variations just follow 
the spectral resolution approximately as
$\Delta\alpha/\alpha\ ({\rm or}\ \Delta\mu/\mu) \propto 1/10{th}$ of the pixel size.
}
\label{fig1}
\end{figure}

This study is aimed to test whether dimensionless physical 
constants are really constants, or whether they vary with space and time.
The latter would imply, at some level,  a
violation of the Einstein equivalence principle (EEP), i.e.,
local position invariance (LPI) and local Lorentz invariance (LLI),
as suggested in a number of unification theories 
(for reviews, see Uzan 2011; Liberati 2013).
In particular, LPI postulates that the outcome of any local nongravitational
experiment is independent of where and when it is performed, i.e.,
that the fundamental physical laws are space-time invariant.
Experimental validation of EEP is one of the most important topics of 
modern physics allowing us to probe the applicability limits of the
Standard Model (SM) of particle physics. 
At the same time, precision limits delivered from such experiments
serve as restrictions for the numerous new theories beyond the SM and can help to distinguish
between them.

Figure~\ref{fig1} demonstrates how upper limits on variations of physical constants
obtained from astronomical spectra just followed the available spectral resolution.
Up to now, no signals have yet been detected in the range of fractional changes
from $\sim 3\times10^{-2}$ to $\sim 3\times10^{-8}$. Thus, any progress in
improving the existing limits can be achieved from observations of narrow
spectral lines involving
higher spectral resolutions to resolve completely their profiles. 
At the moment, the resolution of radio telescopes exceeds that of optical
facilities by order(s) of magnitude; an additional and very attractive property
of microwave radio observations is that some molecular
transitions from this frequency range are extremely sensitive
to the putative variations of the fundamental physical constants
(see a review by Kozlov \& Levshakov 2013).

Flambaum \& Kozlov proposed in 2007 the so-called ammonia method to test the
variability of the electron-to-proton mass ratio, $\mu$. Using this method
for a sample of cold molecular cores from the Milky Way disk we obtained
the following estimate
on the spacial $\mu$-variations (Levshakov et al. 2010a, 2010b):
\dmm\ = $(26\pm1_{\rm stat}\pm3_{\rm sys})\times10^{-9}$ ($1\sigma$ C.L.). However,
further studies revealed significant
instrumental instabilities in the measurements of line radial velocities which were
not accounted for in the above value.
Thus, we performed a new set of observations of the same
targets and with the same instrument (100-m Effelsberg radio telescope) 
in order to get an insight into
this previously unknown systematics. Here we present the recent results.

\section{Observations}

Observations in 2011-2013 targeted a sample of nine
cold ($T_{\rm kin} \sim 10$K)
and dense ($n_{{\rm H}_2} \sim 10^4$ \cmm) starless molecular cores 
located in the Milky Way disk.
The selected clouds are known to have narrow molecular emission lines 
(full width at half maximum, FWHM $< 1$ \kms)
what makes them the most suitable targets to precise measurements of
relative radial velocities (RV).
The following molecular transitions were observed:
\nhhh(1,1) 23.7 GH, \hcIII(2-1) 18.2 GHz, \hcV(9-8) 23.9 GHz, 
\hcVII(16-15) 18.0 GHz, \hcVII(21-20) 23.7 GHz, and \hcVII(23-22) 25.9 GHz.

The source coordinates are taken from Levshakov et al. (2010b, 2013b).
Observations used the Effelsberg 100-m radio telescope as described
in Levshakov et al. (2010a, 2013b). 

In 2011, the measurements were obtained in frequency switching
(FSW) mode using a frequency throw of $\pm2.5$ MHz. The
backend was a fast Fourier transform spectrometer (FFTS),
operated with a bandwidth of 20 MHz, which simultaneously
provided 16\,384 channels for each polarization. 
The resulting channel width was 
0.015 \kms. However, the true velocity resolution is
about 1.6 times coarser.

In 2012-2013, we performed the measurements in the position switching (PSW) mode
with the backend XFFTS (eXtended bandwidth FFTS)
operating with 100\,MHz bandwidth and providing 32\,768 channels
for each polarization. The resulting channel width was 0.039 \kms,
but the true velocity resolution is 1.16 times coarser (Klein et al. 2012).

\begin{figure}[t!]
\resizebox{\hsize}{!}{\includegraphics[clip=true]{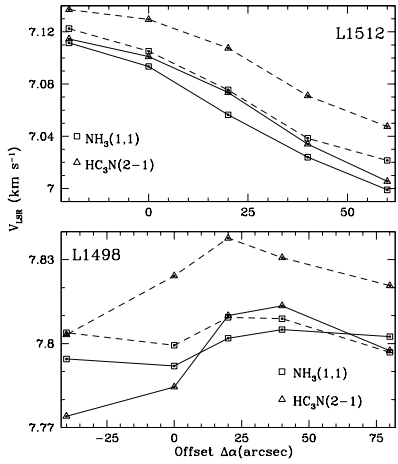}}
\caption{
\footnotesize
The line-of-sight velocities ($V_{\rm LSR}$) of
\nhhh\ $(J,K) = (1,1)$ (squares) and \hcIII\ $J = 2-1$ (triangles)
transitions
at different radial distances along the main diagonal cuts 
towards the molecular cores L1512 and L1498 
measured in 2010 (dashed lines) and in 2011 (solid lines)
at the Effelsberg 100-m radio telescope.
The half-power beam width at 23 GHz is $40''$, the backend
is the fast Fourier transform spectrometer (FFTS) 
with the channel separation $\delta v = 0.015$ \kms.
}
\label{fig2}
\end{figure}

\begin{figure}[t!]
\resizebox{\hsize}{!}{\includegraphics[clip=true]{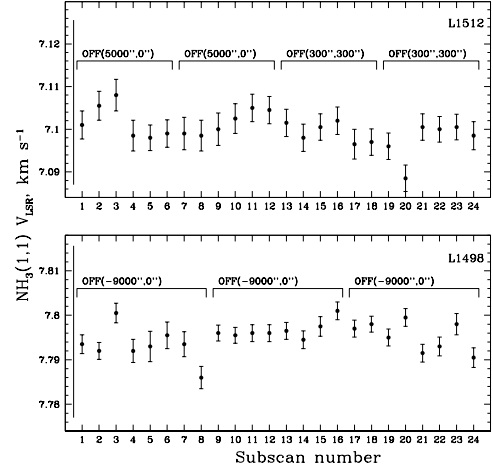}}
\caption{
\footnotesize
The line-of-sight velocities ($V_{\rm LSR}$) of the
\nhhh(1,1) transition (dots with $1\sigma$ error bars)
towards the ammonia peaks in the molecular cores L1512 and L1498 
measured continuously in the position switching mode (PSW) at 
the Effelsberg 100-m radio telescope 
in April, 2012. The exposure time at each point is 150 sec. The PSW offsets are shown in
parentheses.
The backend was an extended fast Fourier transform spectrometer (XFFTS) 
with a channel separation $\delta v = 0.039$ \kms\ (marked by
vertical lines).
The instability of the $V_{\rm LSR}$ measurements of an amplitude $\approx \pm 0.01$ \kms\ 
(i.e., $\approx 1/4th$ of the channel width) is revealed.
}
\label{fig3}
\end{figure}

\section{Results}

To check the reproducibility of the relative RVs of the \nhhh(1,1) and \hcIII(2-1) lines
we re-observed two molecular cores L1512 and L1498 in 2011.
The procedure was the same as in 2010 observations:
cores were mapped at the same offsets and in the same lines.
Namely in the (1,1) inversion transition of \nhhh\ complemented by rotational lines of
other molecular species.
The comparison of radial velocities of \nhhh\ inversion lines, $V_{\rm inv}$,
with radial velocities of rotational transitions, $V_{\rm rot}$,
provides a sensitive limit to the variation of $\mu$ (Flambaum \& Kozlov 2007):
$$
{\Delta \mu}/{\mu} = 0.289(V_{\rm rot} - V_{\rm inv})/c
\approx 0.3\Delta V/c\ ,
$$
where $c$ is the speed of light.

The measured RVs (Levshakov et al. 2010a)
at different radial distances along the main diagonal cuts 
towards L1512 and L1498 are shown in Fig.~\ref{fig2}.
It is seen that the velocity offsets $\Delta V$ exhibit
quite different behavior between 2010 and 2011,
what is probably an effect of unknown systematic errors.

To figure out the source of these errors, we performed in 2012 a set of continuous
observations of L1512 and L1498 targeting their ammonia peaks. 
Observing in PSW mode, we also used different OFF positions to check possible contamination
from an extended background ammonia emission (which was not detected).
The resulting time series are shown in Fig.~\ref{fig3}.
The RV values fluctuate with an amplitude of $\approx \pm 0.01$ \kms, i.e.,
$\approx 1/4th$ of the channel width.

\begin{figure}[t!]
\resizebox{\hsize}{!}{\includegraphics[clip=true]{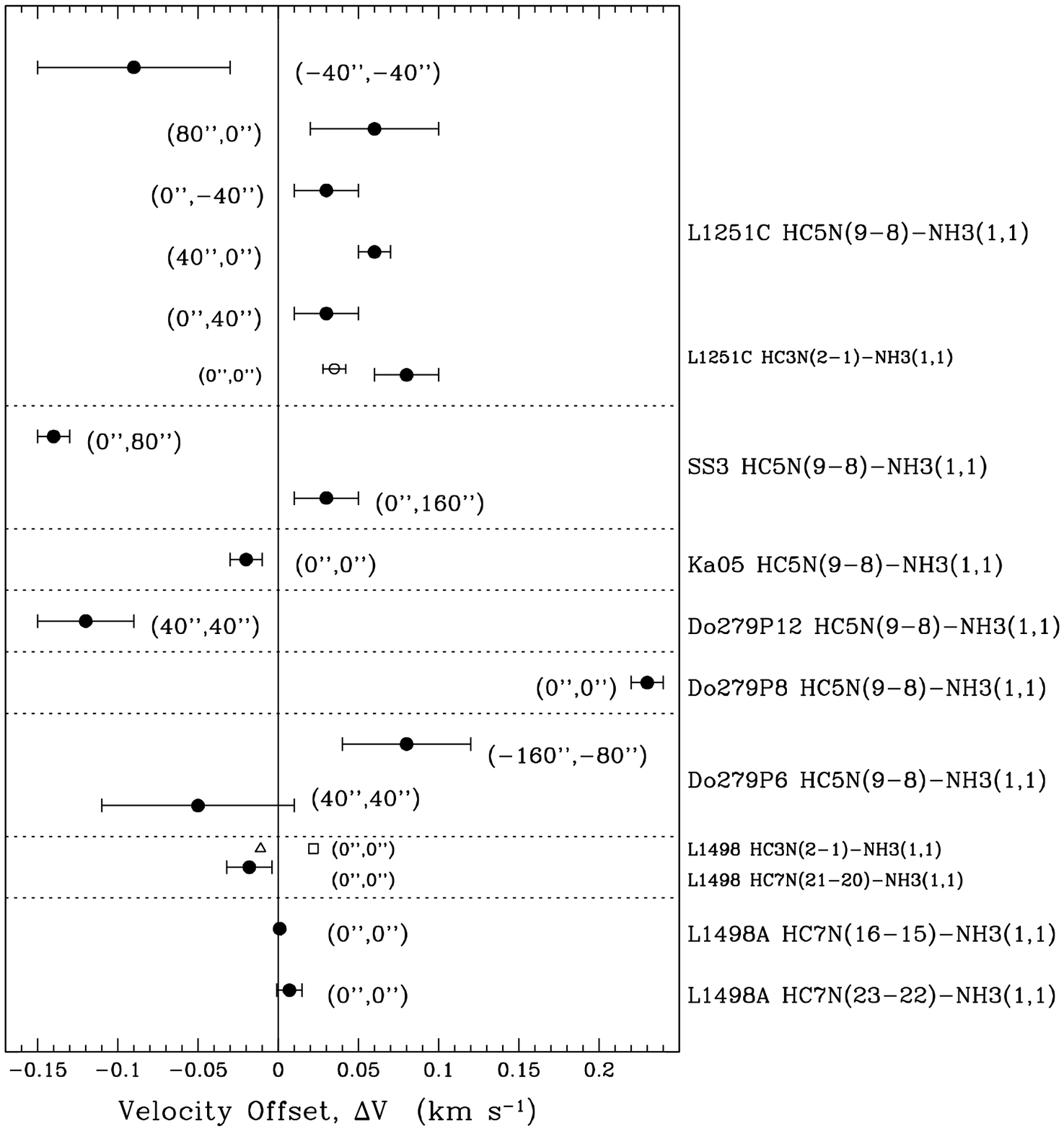}}
\caption{
\footnotesize
Radial velocity differences, $\Delta V$, between rotational 
transitions of different molecules and the 
\nhhh(1,1) line 
for the sources observed at the Effelsberg 100-m telescope (2011-2013).
1$\sigma$ statistical errors are indicated.
In the panel, given in parentheses are the coordinate offsets in arcsec.
Filled circles~-- data from Levshakov et al. (2013a);
open triangle~-- this paper; 
open circle and square~-- Levshakov et al. (2010a, 2010b); 
}
\label{fig4}
\end{figure}

To check whether the sky frequency is identical with the frequency coming out of the
backend we carried out a test with an artificial signal at 22000.78125 MHz.
The synthesizer frequency was accurate to about 1 Hz, and
the frequency scale was found to be accurate to about 32 Hz ($\approx 0.0004$ \kms).

In our observations, the sky frequencies were reset at the onset of each subscan.
Therefore, the longer a subscan the higher
the error caused by Doppler shifts during the exposure time
(e.g., for a 5 min scan, it is about 0.004 \kms\ at Effelsberg latitude).
We corrected some of our observations to account for residual
Doppler shifts. This didn't show a significant change in the results, however.

Another source of errors which can affect the $\Delta\mu/\mu$
estimates with the ammonia method is
the possible segregation of molecules in molecular cores.
Figure~\ref{fig4} shows our measurements in 2013 (filled circles)
of the relative RVs between \nhhh(1,1)  
and other molecules
towards eight molecular cores (L1498A is a gas condensation in L1498,
see Fig.~3 in Kuiper et al. 1996)
at different offsets.
Previous values, obtained in 2009-2011, are marked by open symbols.
A spread of the velocity offsets $\Delta V$ is clearly seen.

Thus, we conclude that noise in the $\Delta V$ values consists of at least two components:
one is due to chemical differentiation and velocity gradients within
the molecular cores, 
possibly being amplified by small variations in the telescope pointing.
Additionally, some scatter in the RVs may be caused by
the different optical depths of the hyperfine structure transitions.
However, all these effects may be random from one observation to another, and,
being averaged over a sample of targets, should be reduced to some extent.
Applied to our sample of $n=18$ independent $\Delta V$ offsets
shown in Fig.~\ref{fig4}, this gives the weighted mean
$\langle \Delta V \rangle = 0.002 \pm 0.015$ \kms\ ($3\sigma$ C.L.).
Being interpreted in terms of
\dmm\ $= (\mu_{\rm obs} - \mu_{\rm lab})/\mu_{\rm lab}$, 
this value of $\langle \Delta V \rangle$
constraints the $\mu$-variation at the level of
$\Delta \mu/\mu  < 2\times10^{-8}$ ($3\sigma$ C.L.),
which is the most stringent limit on the spatial variation of $\mu$
based on radio astronomical observations. 
The same order of magnitude upper limit on \dmm\ was obtained from
independent observations of L1512 and L1498 
at the Medicina 30-m telescope (Levshakov et al. 2013a).

We note in passing that
mapping of the dense molecular cores in different molecular lines shows 
that there is, in general, a good
correlation between \nhhh, N$_2$H$^+$, and \hcIII\ distributions
(Fuller \& Myers 1993; Hotzel et al. 2004; Tafalla et al. 2004; Pagani et al. 2009). 
However, in some clouds \nhhh\ is not traced by \hcIII, as, e.g., in the dark cloud
TMC-1, where peaks of line emission are offset by $7'$ (Olano et al. 1988).
In our case, we observe systematic velocity shifts between \nhhh\ and other species.
This can be expected since C-bearing molecules are usually
distributed in the outer parts of the cores, whereas N-bearing 
molecules trace the inner parts.

\section{Discussion}

The obtained local constraint on the spatial $\mu$-variation, 
$\Delta\mu/\mu < 2\times10^{-8}$ ($3\sigma$ C.L.),
can be used to set limits on changes in $\alpha$. 
This, however, is strongly model dependent.  
For example, within the grand unification model (GUT)
a variation of $\alpha$ would imply considerably larger fractional changes in 
the mass scale of the strong force in QCD, $\Lambda_{\rm QCD}$, 
and in quark and electron masses leading to 
$$
\Delta \mu/\mu \propto R \Delta\alpha/\alpha\ ,
$$
where $R \sim 40$ (e.g., Langacker et al. 2002; Flambaum et al. 2004).
This gives a limit on
$| \Delta\alpha/\alpha | < 10^{-10}$ ($1\sigma$ C.L.).
The value of $R$ is, however, poorly constrained.
Depending on the theory, it varies from $-235$ to +46 (examples are given in Section~5.3.1 in Uzan 2011).
The only way to distinguish between theories is to measure \daa\ and \dmm\ independently.

At higher redshifts, the most stringent limit on cosmological $\mu$-variation
was set at $z = 0.89$, $| \Delta \mu/\mu | < 10^{-7}$  (Bagdonaite et al. 2013).
This would imply that $| \Delta\alpha/\alpha | < 2.5\times10^{-9}$ ($1\sigma$ C.L.)
at epoch $7\times10^9$ yr, which means in turn that 
$| \dot{\alpha}/\alpha | < 4\times10^{-19}$ yr$^{-1}$. 
At very high redshift, $z = 5.2$ (epoch 12.9 Gyr),
the current limit is $| \Delta\alpha/\alpha | < 8\times10^{-6}$ ($1\sigma$ C.L.)
corresponding to $|\dot{\alpha}/\alpha| < 6\times10^{-16}$ yr$^{-1}$
(Levshakov et al. 2012).
We note that most stringent limits set by the Oklo fossil reactor 
and by terrestrial atomic clock experiments are, respectively,
$|\dot{\alpha}/\alpha| < 5\times10^{-17}$ yr$^{-1}$ (Uzan 2011), and
$|\dot{\alpha}/\alpha| < 4\times10^{-17}$ yr$^{-1}$ (Rosenband et al. 2008).
Thus, despite many efforts, the space-time variations in $\mu$ and $\alpha$ have never
been detected either in laboratory or astronomical experiments.

It only remains to hope that significant
improvements in radio 
astronomical observations will allow us to probe
variations of $\mu$ at levels of
$\Delta \mu/\mu \sim 10^{-9}-10^{-10}$,
leading to even more stringent results and
eventually to the detection of a real variation.

\section{ Conclusions}

We have used the Effelsberg 100-m telescope to observe the
\nhhh(1,1) 23.7 GHz, 
\hcIII(2-1) 18.2 GHz, \hcV(9-8) 23.9 GHz, \hcVII(16-15) 18.0 GHz, \hcVII(21-20) 23.7 GHz, and \hcVII(23-22) 25.9 GHz 
spectral lines in high-density molecular cores devoid of associated IR sources.
The results obtained are as follow.
\begin{enumerate}
\item[1.] In order to test the reproducibility of the measurements of the relative
radial velocities between the
\nhhh(1,1) and \hcIII(2-1) transitions observed towards dark molecular cores in 2009-2010
at the Effelsberg 100-m telescope, we re-observed two clouds L1512 and L1498 and revealed
discrepancy between the $V_{lsr}({\rm HC}_3{\rm N}) - V_{lsr}({\rm NH}_3)$
values which is as high as the channel width, $\Delta V \la 0.02$ \kms.
\item[2.] Continuous observations of L1512 and L1498 in 2012 at a fixed position towards
the ammonia peaks showed that the measured radial velocity $V_{lsr}({\rm NH}_3)$ fluctuates
during the exposure time of 2 hours with an amplitude $\approx \pm 0.01$ \kms, i.e.,
with approximately $1/4th$ of the channel width.
\item[3.] Tests with the synthesizer frequency at 
2000.78125 MHz showed that the sky frequency is accurate to about 32 Hz, i.e.,
$\approx 0.0004$ \kms.
\item[4.] Taking into account the revealed errors and averaging relative velocities over
a sample of eight molecular cores ($n=18$ independent $\Delta V$ values)
observed in 2013, we find a null offset
$\langle \Delta V \rangle = 0.002 \pm 0.015$ \kms\ ($3\sigma$ C.L.) 
between the rotational and inversion transitions of the above mentioned molecules.
\item[5.] If this offset is interpreted in terms of
\dmm\ $= (\mu_{\rm obs} - \mu_{\rm lab})/\mu_{\rm lab}$, then 
the spatial $\mu$-variation is constrained at the level of
$\Delta \mu/\mu  < 2\times10^{-8}$ ($3\sigma$ C.L.), 
that is the strictest limit for the validity of the LPI principle
based on radio astronomical observations. 
\end{enumerate}

\begin{acknowledgements}
We thank the staff of the Effelsberg 100-m telescope for the assistance in observations and
acknowledge the help of Benjamin Winkel in preliminary data reduction.
SAL's work is supported by the grant DFG
Sonderforschungsbereich SFB 676 Teilprojekt C4, 
and in part
by Research Program OFN-17 of the Division of Physics, 
Russian Academy of Sciences.
\end{acknowledgements}

\bibliographystyle{aa}

\begin{thebibliography}{}


\bibitem[{Agafonova et al. (2011)}]{a11} 
Agafonova, I. I., Molaro, P., Levshakov, S. A., \& Hou, J. L.
2011, A\&A, 529, A28

\bibitem[{Bagdonaite et al. (2013)}]{b13}
Bagdonaite, J., Jansen, P., Henkel, C., Bethlem, H. L., Menten, K. M., \& Ubachs, W. 
2013, Science, 339, 46

\bibitem[{Bahcall et al. (1967)}]{bss} 
Bahcall, J. N., Sargent, W. L. W., \& Schmidt, M. 
1967, ApJ, 149, L11

\bibitem[{Chand et al. (2005)}]{c05} 
Chand, H., Petitjean, P., Srianand, R., \& Aracil, B. 
2005, A\&A, 430, 47

\bibitem[{Flambaum \& Kozlov (2007)}]{fk} 
Flambaum, V. V. \& Kozlov, M. G. 2007, PhRvL, 98, 240801

\bibitem[{Flambaum et al. (2004)}]{f04} 
Flambaum, V. V., Leinweber, D. B., Thomas, A. W., \& Young, R. D. 
2004, PhRvD, 69, 115006

\bibitem[{Fuller \& Myers (1993)}]{fm} 
Fuller, G. A., \& Myers, P. C. 1993, ApJ, 418, 273

\bibitem[{Hotzel et al. (20040)}]{hot}
Hotzel, S., Harju, J., \& Walmsley, C. M. 
2004, A\&A, 415, 1065 

\bibitem[{Klein et al. (2012)}]{kl12} 
Klein, B., Hochg\"urtel, S., Kr\"amer, I., Bell, A., Meyer, K., \& G\"usten, R. 
2012, A\&A, 542, L3

\bibitem[{Kozlov \& Levshakov (2013)}]{kl} 
Kozlov, M. G., \& Levshakov, S. A. 2013, Ann. Phys., 525, 452

\bibitem{}
Kuiper, T. B. H., Langer, W. D., \& Velusamy, T. 
1996, 468, 761

\bibitem[{Langacker et al. (2002)}]{la02} 
Langacker, P., Segr\'e, G., \& Strassler, M. J.
2002, Phys. Lett. B, 528, 121 

\bibitem[{Levshakov 1992}]{lev92} 
Levshakov, S. A. 1992, in High Resolution Spectroscopy with the VLT,
ed. M.-H. Ulrich (ESO: Garching/Munchen), p.139

\bibitem[{Levshakov et al. (2010a)}]{lev10a} 
Levshakov, S. A., Lapinov, A. V., Henkel, C., Molaro, P., Reimers, D., Kozlov, M. G., Agafonova, I. I. 
2010a, A\&A, 524, A32

\bibitem[{Levshakov et al. (2010b)}]{lev10b} 
Levshakov, S. A., Molaro, P., Lapinov, A. V., Reimers, D., Henkel, C., \& Sakai, T. 
2010b, A\&A, 512, A44

\bibitem{}Levshakov, S. A., Combes, F., Boone, F.,
Agafonova, I. I., Reimers, D., \& Kozlov, M. G. 2012, A\&A, 540, L9

\bibitem[{Levshakov et al. (2013a)}]{lev13a} 
Levshakov, S. A., Reimers, D., Henkel, C., Winkel, B., Mignano, A., Centuri\'on, M., \& Molaro, P. 
2013a, A\&A, submitted  

\bibitem[{Levshakov et al. (2013b)}]{lev13b} 
Levshakov, S. A., Henkel, C., Reimers, D., Wang, M., Mao, R., Wang, H., \& Xu, Y.
et al. 2013b, A\&A, 553, A58

\bibitem[{Liberati (2013)}]{li13} 
Liberati, S. 2013, CQGra, 30, 133001

\bibitem[{Molaro et al. (2013)}]{m13} 
Molaro, P., Centuri\'on, M., Whitmore, J. B.,  et al. 
2013, A\&A, 555, A68

\bibitem[{Murphy et al. (2001)}]{m01} 
Murphy, M. T., Webb, J. K., Flambaum, V. V., Prochaska, J. X., \& Wolfe, A. M. 
2001, MNRAS, 327, 1237

\bibitem[{Olano et al. (1988)}]{o88} 
Olano, C. A., Walmsley, C. M., \& Wilson, T. L.
1988, A\&A, 196, 194

\bibitem[{Pagani et al. (2009)}]{p09} 
Pagani, L., Daniel, F., \& Dubernet, M.-L. 
2009, A\&A, 494, 719

\bibitem[{Quast et al. (2004)}]{q04} 
Quast, R., Reimers, D., \& Levshakov, S. A.  
2004, A\&A, 415, L7

\bibitem[{Rosenband et al. (2008)}]{ro08} 
Rosenband, T., Hume, D. B., Schmidt, P. O., et al.
2008, Science, 319, 1808

\bibitem[{Savedoff 1956}]{sav}
Savedoff, M. P. 1956, Nature, 178, 688

\bibitem[{Srianand et al. (2007)}]{s07} 
Srianand, R., Chand, H., Petitjean, P., \& Aracil, B.
2007, PhRvL, 99, 239002

\bibitem[{Tafalla et al. (2004)}]{t04} 
Tafalla, M., Myers, P. C., Caselli, P., \& Walmsley, C. M. 
2004, A\&A, 416, 191

\bibitem[{Uzan (2011)}]{u11}
Uzan, J.-P. 2011, Living Rev. Relativity, 14, 2

\bibitem[{Varshalovich \& Potekhin (1995)}]{vp} 
Varshalovich, D. A., \& Potekhin, A. Y. 1995, SSRv, 74, 259

\bibitem[{Varshalovich et al. (1996)}]{v96} 
Varshalovich, D. A., Panchuk, V. E., \& Ivanchik, A. V. 
1996, AstrL, 22, 6

\bibitem[{Webb et al. (1999)}]{w99} 
Webb, J. K., Flambaum, V. V., Churchill, C. W., Drinkwater, M. J., \& Barrow, J. D. 
1999, PhRvL, 82, 884


\end{thebibliography}

\end{document}